\documentclass{elsart5p}

\usepackage{graphics}
\usepackage{graphicx}
\usepackage{amssymb}
\begin{document}

\begin{frontmatter}



\title{Quantum phase transitions beyond the dilute Bose gas limit }

\author{Valeri N. Kotov},
%\ead{kotov@bu.edu}
\author{D. X. Yao},
\author{A. H. Castro Neto},
\author{D. K. Campbell}

\address{Department of Physics, Boston University, 590 Commonwealth Avenue,
Boston, MA 02215}

%\corauth[Kotov]{Corresponding author.} 

\begin{abstract}
% Text of abstract
We  study a Heisenberg S=1/2   ring-exchange antiferromagnet
 which exhibits a quantum phase transition from a spontaneously dimerized
 (valence bond solid) phase to a magnetically ordered (N\'eel) phase.
 We argue that the quantum transition is of unconventional nature;
 both   singlet and triplet modes of high density condense as
the transition is approached from the dimer side, signaling restoration of lattice symmetry.
 These  features are consistent with ``deconfined quantum criticality",
 of which the present model is believed to be the only example so far.
\end{abstract}


\end{frontmatter}

\begin{figure}[!ht]
\begin{center}
\includegraphics[angle=0,width=0.44\textwidth]{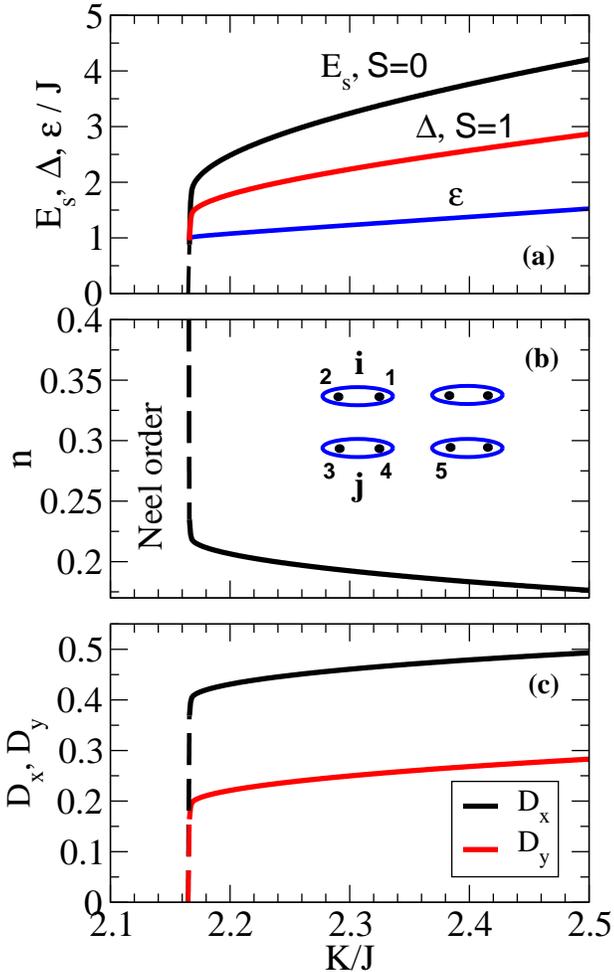}
\end{center}
\caption{(a.) Lowest triplet ($\Delta$) and singlet ($E_{s}$) energy gaps
 near the QCP. The singlet binding energy $\epsilon = 2\Delta - E_s$. 
(b.) Triplon quasiparticle density $n$. (c.) Dimer order parameters.} \label{Fig1}
\end{figure}

% main text
Recent work on quantum critical phenomena has revealed the  fascinating
 possibility of ``deconfined criticality", where spinon deconfinement
 occurs at a quantum critical point (QCP) separating a
 valence bond solid (VBS) and magnetically ordered (N\'eel) phase \cite{Senthil}.   
 In order for this exotic scenario to take place, the magnetic and VBS orders
 must not coexist; otherwise the spinons remain permanently
 ``paired" in S=1 quasiparticles \cite{Senthil,Levin}. The intense search for
 a microscopic Heisenberg model that exhibits this novel criticality has led
 to a solid possibility: a S=1/2 model with four-spin ring exchanges \cite{Sandvik}.
 The Hamiltonian of this two-dimensional, square-lattice model reads
\begin{equation}
\label{model}
H = J \sum_{\langle i,j \rangle} {\bf{S}}_{i}.{\bf{S}}_{j}
-K \sum_{i,j,k,l} ({\bf{S}}_{i}\cdot{\bf{S}}_{j})({\bf{S}}_{k}\cdot{\bf{S}}_{l}),
\end{equation}
where the couplings are antiferromagnetic, $J,K > 0$. The four-spin term acts on a given
 plaquette as 
$-K \{ ({\bf{S}}_{1}\cdot{\bf{S}}_{2})({\bf{S}}_{3}\cdot{\bf{S}}_{4})
+  ({\bf{S}}_{2}\cdot{\bf{S}}_{3})({\bf{S}}_{1}\cdot{\bf{S}}_{4}) \}$,
where the spins are numbered as in Fig.~\ref{Fig1}(b). Summation  is performed
over all plaquettes, so that $H$ does not break the square lattices symmetries.
 The quantum Monte Carlo study of Ref.~\cite{Sandvik} found a critical
 coupling $(K/J)_{c} \approx 2$, separating a VBS phase  ($K > K_{c}$) and
 a N\'eel phase ($K < K_{c}$). The spontaneous dimerization in the VBS state 
 was suggested to be of the columnar dimer type, as  in  Fig.~\ref{Fig1}(b).
 
  In the present work we study the model (\ref{model}) by approaching the QCP from
 the dimerized phase. In this case it is natural to represent the spins in terms of
 ``bond triplons" $t_{{\bf i}{\alpha}}^{\dagger}, \alpha=x,y,z$  \cite{Sachdev} which
 create  S=1 triplet excitations  on a given dimer ${\bf i}$ (see  Fig.~\ref{Fig1}(b)).
 These quasiparticles are hard-core, meaning that two triplons cannot be created
 on the same site (dimer). The analysis of the effective Hamiltonian $H_{\mbox{eff}}[t_{\alpha}]$,
 written in terms of the triplon
 operators, leads to the spectrum $\omega ({\bf k})$ 
  which determines the location of the QCP where the magnetically
 ordered state emerges.  

 Even though the representation of the original spin Hamiltonian 
 in terms of $H_{\mbox{eff}}[t_{\alpha}]$ is exact, in practice one has to deal
 with the various interactions between the triplons in an approximate way.
 These interactions include both the kinematic hard-core constraint
 and the various dynamic triplon-triplon scattering vertices.
 For the model (\ref{model})  up to 8-point vertices appear.
 The success of perturbation theory in treating
 such interactions is determined by the quasiparticle density $n$ which
 it typically low (microscopically the density is related to the 
 quantum fluctuations).
 In quantum spin models involving explicit dimerization,
 i.e. situations where some of the exchange coupling are stronger than the
 others (such as spin ladders, coupled ladders, bi-layer models, etc.),
 calculations based on the triplon representation are reliable even at the
 mean-field level \cite{Sachdev}. Improvements beyond mean-field theory
 can be made by resummation of selected diagrams (at lowest
 order in the density) as typically done in the case of a dilute Bose gas
  \cite{Kotov1}. This leads to accuracy within 10 percent,
 corresponding roughly to the low  quasiparticle density $n \lesssim 0.1$.

It is clear that since for the model (\ref{model})  the VBS order is
 spontaneous and is expected to disappear at the QCP,
 the triplons are not necessarily in the dilute limit, especially
 around the QCP. This makes the analysis quite involved and the results
 somewhat dependent on the level of approximation.  
 Moreover, the destruction of the dimer order at the QCP is expected
 to lead to vanishing of a singlet energy scale $E_s$ (in addition
 to the vanishing of the triplon gap). Here we present  our main results
 while the technical details will be available elsewhere \cite{Kotov2}. 

 We have found that the quantum fluctuations in the dimer background
 are strong, which is evident from the  fact that the four-spin K term
 in (\ref{model}) favors equally ``vertical" dimer ladders (as in   Fig.~\ref{Fig1}(b))
 and horizontal ones, obtained by a 90 degree rotation. 
 Since the triplons are constructed on a fixed dimer pattern, such a 
 tendency towards restoration of rotational symmetry is related 
 to the  tendency of the triplons to form ``pairs" with total spin S=0,
 in essence reflecting ``plaquettization". Within our approach this is
 a non-perturbative effect. 
 We denote the energy of such a two-triplon singlet excited  state 
 by $E_s$ (it is easy to see that the lowest singlet branch
 also carries zero total momentum), and its wave-function is of the form
 $|\Psi \rangle_{s} = \sum_{\alpha,{\bf q}_y} \Psi({\bf q}) \
 t_{\alpha, {\bf q}}^{\dagger} t_{\alpha, {-\bf q}}^{\dagger}|0\rangle$.
The pairing is most effective along the  vertical (``y") direction since
 in this case the plaquette interaction acts fully on two dimers,
whereas pairing involving four triplons is energetically much less  favorable. 
 The excitation  energy $\omega ({\bf k})$ 
of the one-triplon ($t_{\alpha, {\bf k}}^{\dagger}|0\rangle$) spin S=1 state  
has a minimum at the N\'eel ordering  wave-vector ${\bf k}_{AF}\equiv (\pi,\pi)$;
 we define the gap as $\Delta \equiv \omega ({\bf k}_{AF})$.
 The evolution of $\Delta$ and $E_s$ as a function of the coupling
$K/J$ is shown in Figure 1(a).
We find that the  quantum phase transition to the N\'eel phase takes place at $(K/J)_{c} \approx 2.16$,
 where $\Delta \rightarrow 0$, in good agreement with the Monte Carlo
 result \cite{Sandvik}. In addition,  the  singlet energy scale $E_s$ also vanishes
at the QCP, reflecting the destruction of the dimer order. The direct evaluation of the two dimer order
 parameters $D_{x} = |\langle {\bf S}_3 \cdot {\bf S}_4 \rangle -
 \langle {\bf S}_5 \cdot {\bf S}_4 \rangle |$ and   
$D_{y} = |\langle {\bf S}_3 \cdot {\bf S}_4 \rangle - \langle {\bf S}_1 \cdot {\bf S}_4 \rangle|$,
 where spins are numbered as in Fig.~\ref{Fig1}(b), indeed shows that they exhibit 
 a tendency to vanish at  the QCP. $D_{x},D_{y}$ are plotted in Figure 1(c). 

The unconventional merger of singlet and triplet modes at the QCP is accompanied
 however by an increase of the triplon quasiparticle density
$n = \sum_{\alpha} \langle t_{{\bf i}\alpha}^{\dagger} t_{{\bf i}\alpha}\rangle$,
 as shown in  Figure 1(b). This  signals a breakdown of our quasiparticle description
 based on a picture of strongly-interacting, but dilute Bose gas of triplons.
 Within  the  effective bond triplon theory such 
 a breakdown is in a way a necessity if  the dimer order  vanishes at the QCP
 and the lattice symmetry is restored. In our  view the relatively sharp variation
 of all quantities near the QCP is an artifact of our self-consistent scheme
 based on the assumption $n \ll 1$; we thus cannot explore reliably the variation
 of the energy scales as $(K/J) \rightarrow (K/J)_c$. A consistent classification
 of diagrams beyond the conventionally used ``ladder" ones  
 in the dilute limit \cite{Kotov1}  is in itself an  unsurmountable task.

 The diverging length scale $L_s \sim 1/E_s$ is not the spinon deconfinement
scale \cite{Senthil} since the state $|\Psi \rangle_{s}$ defined above does not
 carry any topological quantum numbers. However $L_s$ is related to the destruction
 of the dimer order and thus is expected  to implicitly reflect deconfinement; $L_s$ is 
indeed the most natural scale to introduce within the effective triplon theory. The presence
 of $L_s$ in addition to the  diverging magnetic scale $L_M \sim 1/\Delta$ at the QCP 
 signals a tendency towards restoration of lattice symmetry and thus supports
 the notion that the Heisenberg model (\ref{model}) exhibits ``deconfined criticality" \cite{Sandvik}.

This work was supported  by NSF grant DMR-0343790 (A.H.C.N.) and
 by Boston University (V.N.K., D.X.Y, and D.K.C.).

\end{document}